\documentclass[prl,twocolumn,tightenlines,showpacs,floatfix,nofootinbib]{revtex4}
\usepackage{multirow}
\usepackage{amsmath}
\usepackage{bm}
\usepackage{dcolumn}

\newcommand{\tref}[1]{Table~\ref{#1}}

\begin{document}
\title{Long range interactions of ytterbium in mixed quantum gases}

\author{S.~G.~Porsev$^{1,2}$}
\author{M.~S.~Safronova$^{1,3}$}
\author{A. Derevianko$^4$}
\author{Charles~W.~Clark$^3$}
\affiliation{ $^1$Department of Physics and Astronomy, University of Delaware,
    Newark, Delaware 19716\\
$^2$Petersburg Nuclear Physics Institute, Gatchina, Leningrad District, 188300, Russia \\
$^3$Joint Quantum Institute, National Institute of Standards and Technology and the \\
University of Maryland, Gaithersburg, Maryland, 20899\\
$^4$Physics Department, University of Nevada, Reno, Nevada 89557}
\date{\today}

\begin{abstract}
We present methods for accurate evaluation of van der Waals coefficients of dimers
with excited atoms that have a strong decay channel. We calculate
$C_6$ coefficients for the  Yb--Yb $^1\!S_0+\,^3\!P^o_{0,1}$, $^3\!P^o_0+\,^3\!P^o_{0}$ and Yb--Rb
 $^3\!P_1^o+\,5s \hspace {3 pt} ^2\!S_{1/2}$, $^1\!S_0+\, 5p \hspace {3 pt} ^2\!P_{1/2}^o $ dimers and
$C_8$ coefficients for the Yb--Yb $^1\!S_0+\,^1\!S_0$, $^1\!S_0+\,^3\!P^o_1$ and
Yb--Rb $^1\!S_0+\,5s~^2\!S_{1/2}$  dimers.
We evaluate uncertainties of all properties.
% Our result $C_8=1.88(9) \times 10^{5}$~a.u. for Yb-Yb $^1\!S_0+\,^1\!S_0$ dimer
% is in excellent agreement with the value derived from the experiment $1.9(5)\times 10^5$~a.u.
Our $C_8$ for Yb--Yb $^1\!S_0+\,^1\!S_0$ agrees with the recent experimental value, and is 8 times more accurate.
\end{abstract}
\maketitle

%\paragraph{Introduction}

Ytterbium (Yb), with atomic number $Z = 70$ and ground-state configuration [Xe]$4f^{14}6s^2$, has emerged as a preferred candidate for the study of quantum gases~\cite{TakKomHon04}, optical atomic clocks~\cite{Ybclock}, quantum information processing~\cite{GorReyDal09}, and studies of fundamental symmetries~\cite{PNC1}.
With five bosonic and two fermionic stable isotopes in natural abundance, the $^1\!S_0$ ground state, the long-lived metastable $6s6p\,\, ^3\!P_0^o$ state, and transitions at convenient wavelengths for laser cooling and trapping, Yb stands out as a prospect for use in quantum gas mixtures ~\cite{BorCiuJul09,TakKomHon04,TojKitEno06,EnoKitToj08,KitEnoKas08,YamTaiSug13,TakSaiTak12,NemBauMun09,MunBruMad11,BauMunGor11}.
The best limit for electron electric-dipole (EDM) moment that constrained various extensions of the standard model of electroweak interactions was obtained using YbF molecule ~\cite{HudKarSma11}.
Yb-Yb and Yb--Rb dimers studied in this work are proposed for quantum simulation applications and search for electron EDM~\cite{MicBreZol06,ReiJulDeu09,MeyBoh09,SanOdoJav11}.

This brings urgency to understanding the collisional interactions of Yb, both among its various isotopes and with other  gases.  In particular, interactions of Yb atoms in the $^1\!S_0$ and $^3\!P_0^o$ clock states on one hand limit accuracy of optical lattice clocks~\cite{LudLemShe11,BisLinSwa11} and on the other hand can be harnessed to engineer  metrologically significant entanglement of clock atoms~\cite{YeReyScience2013forthcoming}. Characterization of these interactions for Yb is crucial for evaluating  the feasibility of  applying such ideas  to Yb lattice clocks. As for the interactions of Yb with alkali atoms,  their characterization is crucial  for selecting efficient pathways for assembling Yb-alkali molecules via photo- or magneto-association and
STIRAP techniques~\cite{MunBruMad11}. Unlike
ultracold alkali dimer molecules~\cite{NiOspMir08} with largely diamagnetic ground states, the alkali-Yb dimers possess  unpaired electron spin, and thereby can be controlled by both electric and magnetic fields. This added control enlarges the class of many-body Hamiltonians that can be  simulated with ultracold molecules~\cite{MicBreZol06}.

Accurate analysis of these interactions is a challenge for any atom, and usually requires both experimental and theoretical efforts.  Key to both approaches is the determination of the van der Waals coefficients that express the long-range interactions between the two atoms.  Here we report the results of a new approach for calculating the $C_6$ and $C_8$ coefficients of the interaction between ground- and excited-state Yb atoms, and between Rb and Yb atoms.

In previous work ~\cite{SafPorCla12}, we evaluated the $C_6$ coefficient for Yb--Yb
$^1\!S_0 +\,^1\!S_0$ and found it to be $C_6=1929(39)$, in excellent agreement with the experimental
result $C_6=1932(35)$~\cite{KitEnoKas08}. However, our previous approach cannot be directly applied to
the calculation of the van der Waals coefficients for the  Yb--Yb $^1\!S_0+\, ^3\!P^o_1$ and Yb--Rb $^3\!P^o_1 +\, 5s \hspace {3 pt} ^2\!S_{1/2}$ dimers owing to
the presence of the $^3\!P^o_1 -\,^1\!S_0$ radiative decay channel. This is also the case for the  Yb--Rb $^1\!S_0 +\, 5p \hspace {3 pt} ^2\!P^o_{1/2}$ dimer.

Now we report the development of methods for accurate evaluation of the van der Waals coefficients of dimers involving excited-state atoms with strong decay channels to the ground state. These methods can be used for evaluation of van der Waals coefficients in a variety of systems. Here we apply these to cases of current experimental interest. Specifically, in this work we calculated two sets of the van der Waals coefficients: (1)
$C_6$  for the $^1\!S_0+\,^3\!P^o_{0,1}$, $^3\!P^o_0+\,^3\!P^o_{0}$  and
$C_8$  for the $^1\!S_0+\,^1\!S_0$, $^1\!S_0+\,^3\!P^o_1$ Yb-Yb dimers; (2)~$C_6$ for
the $^3\!P_1^o+\,5s \hspace {3 pt} ^2\!S_{1/2}$, $^1\!S_0+\, 5p \hspace {3 pt} ^2\!P_{1/2}^o $  and
$C_8$ for the $^1\!S_0+\,5s~^2\!S_{1/2}$ Yb--Rb  dimers.
Unless stated otherwise, throughout this paper we use atomic units (a.u.);
the numerical values of the elementary
 charge, $|e|$, the reduced Planck constant, $\hbar = h/2
\pi$, and the electron mass, $m_e$, are set equal to 1.
%The atomic unit for polarizability can be converted to SI units via
%$\alpha/h$~[Hz/(V/m)$^2$]=2.48832$\times10^{-8}\alpha$~(a.u.), where the conversion %coefficient is $4\pi \epsilon_0
%a^3_0/h$, $a_0$ is the Bohr radius and $\epsilon_0$ is the dielectric constant.

The $C_6$ coefficients can be obtained from experimental data by determining an asymptotic potential  $-C_6/R^6$ that reproduces experimental values of dimer rovibrational levels, where $R$ is internuclear coordinate.
One of the main uncertainties in this approach is due to the presence of the next order term, $-C_8/R^8$, which is not accurately known for any of the dimers involving Yb.  For example, the present experimental value of $C_8$ of
Yb--Yb
$^1\!S_0+\,^1\!S_0$ is $C_8=1.9(5) \times 10^{5}$~a.u.~\cite{KitEnoKas08}, which has an uncertainty of 25\%. The result that we obtained from the calculations in the present work is
$C_8=1.88(6) \times 10^{5}$~a.u., which is in excellent agreement with the experimental
value, and reduces the uncertainty 3\%.
By constraining the likely range of $C_8$ values, our calculations  may  significantly improve the accuracy of $C_6$ values determined from experimental  photoassociation spectroscopy.

% ===========================
%\paragraph{General formalism for Yb--Yb dimer}
% ===========================
We investigate here the molecular potentials asymptotically connecting to
the $|A\rangle + |B\rangle$ atomic states with fixed total angular
momenta $J_A$ and $J_B$ and their projections $M_A$ and $M_B$. The
combined wave function constructed from these states is
$
|M_A,M_B;\,\Omega \rangle =|J_A M_A\rangle_{\mathrm{I}}\,
|J_B M_B\rangle_{\mathrm{II}},
$
where the index I (II) describes the wave function located on the center I
(II) and $\Omega = M_A+M_B$ is the sum of projections on the internuclear axis.
The molecular wave function $\Psi^{g/u}_\Omega$ of Yb--Yb dimer with $|A\rangle = |^1\!S_0\rangle$ and
$|B\rangle = |^3\!P^o_{0,1}\rangle$,
formed as a linear combination of the wave functions given above,
%--------------------------------------------------------------------------------
\begin{equation}
\Psi^p_\Omega = \frac{1}{\sqrt{2}} ( |A \rangle_{\mathrm{I}}
\,|B \rangle_{\mathrm{II}} + (-1)^p |B \rangle_{\mathrm{I}} |A \rangle_{\mathrm{II}} ),
\end{equation}
%--------------------------------------------------------------------------------
possesses a definite gerade/ungerade symmetry and definite
quantum number $\Omega$. We put $p=0$ for ungerade symmetry and $p=1$ for gerade symmetry, taking into account that
$A$ and $B$ are the opposite parity states.

The dispersion potential describing long-range interaction of two
atoms can be written in the form
%--------------------------------------------------------------------------------
\begin{equation}
U(R) \approx -\frac{C_3}{R^{3}}-\frac{C_6}{R^{6}}-\frac{C_8}{R^{8}}  ,
\label{UR}
\end{equation}%
%--------------------------------------------------------------------------------
where $C_3$ is the coefficient of the dipole-dipole interaction in first-order
of the perturbation theory and $C_6$ and $C_8$ are the coefficients of the dipole-dipole and
dipole-quadrupole interactions in the second-order perturbation theory, respectively.
If $A$ and $B$ are spherically symmetric states with no downward transitions,
the $C_6$ and $C_8$ coefficients for the $A+B$ dimers are (see, e.g.,~\cite{PatTan97})
%===================================================================
\begin{eqnarray}
C^{AB}_6 &=&\frac{3}{\pi}\, \int_0^\infty\, \alpha_1^A(i \omega)\, \alpha_1^B(i \omega)\, d\omega , \nonumber \\
C^{AB}_8 &=&  \frac{15}{2\pi}\, \int_0^\infty\, \alpha_1^A(i \omega)\,
\alpha_2^B(i \omega)\, d\omega \nonumber \\&+& \frac{15}{2\pi}\, \int_0^\infty\, \alpha_2^A(i \omega)\,
\alpha_1^B(i \omega)\, d\omega ,
\label{vdW}
\end{eqnarray}
%===================================================================
where $\alpha_1(i \omega)$ and $\alpha_2(i \omega)$ are dipole and quadrupole
dynamic polarizabilities at an imaginary frequency.
%===================================================================

For the $^1\!S_0 +\, ^3\!P_1^o$ dimer, the expressions for $C_6$ and $C_8$ derived
 in the present work are substantially more complicated due to different  angular dependence
and the $^3\!P^o_1 \rightarrow \, ^1\!S_0$ decay channel to the ground state.
A derivation of respective formulas is rather lengthy and will be presented in
a subsequent detailed paper. The $C_{6}$ coefficient in this case is given by
$
C_{6}(\Omega _{p})=\sum_{J=0}^{J=2}A_{J}(\Omega )X_{J},
$
where $A_J$ are the angular factors, $\Omega=0,1$, and
the quantities $X_J$ ($J=0,1,2$) are
\begin{eqnarray}
&&X_J = \frac{27}{2\pi} \int_0^\infty \alpha_1^A(i\omega) \,
\alpha^B_{1 J}(i\omega) \, d\omega \\
&+&\! \delta_{J,0} \left[ 2D^{2}\sum_{n\neq B}\frac{(E_n-E_A)\,
|\langle n||d||A \rangle |^{2}}{(E_{n}-E_A)^{2}-\omega_0^2}+\frac{D^{4}}{2\,\omega_0} \right]\!.\nonumber
\label{delX0}
\end{eqnarray}%
Here $A\equiv\, ^1\!S_0$, $B\equiv\, ^3\!P^o_1$, $D\equiv |\langle ^{3}\!P_{1}^{o}||d||^{1}\!S_{0}\rangle|$,
$\omega _{0}\equiv E_{\,^{3}\!P_{1}^{o}}-E_{\,^{1}\!S_{0}}$, and $\alpha^B_{1 J}$ designates a part of the scalar
dipole $^3\!P^o_1$ polarizability, where the sum ranges over intermediate states $n$ with fixed total angular momentum $J_n=J$.
The expression for the $C_{8}$ coefficient for the Yb--Yb $^1\!S_0 +\, ^3\!P_1^o$  dimer
is considerably more complicated.
%\begin{equation}
%C_8(\Omega_p)= \sum_{k=1}^4 \sum_{J_a=1}^2 \sum_{J_b=0}^3
%A_k^{J_a J_b}(\Omega_p) X_k^{J_a J_b} .
%\label{C8}
%\end{equation}
The detailed final expressions for the $C_6$ and $C_8$ coefficients are given
in the Supplemental Material~\cite{SupMat}.

% ===============================
%\paragraph{Method of calculation}
To estimate the uncertainty of our results we carried out the Yb calculations using two different methods.
The first method combines configuration interaction (CI) with many-body perturbation theory
(MBPT)~\cite{DzuFlaKoz96b}. In the second, more accurate method, CI is combined with the coupled-cluster
all-order approach (CI+all-order) that treats both core and valence correlations to
all orders~\cite{Koz04,SafKozJoh09,SafKozCla11}.

The point of departure for both methods is a solution of the Dirac-Fock (DF) equations,
$
\hat H_0\, \psi_c = \varepsilon_c \,\psi_c,
$
where $H_0$ is the relativistic DF Hamiltonian~\cite{DzuFlaKoz96b,SafKozJoh09} and $\psi_c$ and $\varepsilon_c$ are
single-electron wave functions and energies. The details of the Yb basis set and the CI space construction were
given in \cite{SafPorCla12}.
The wave functions and the low-lying energy levels are determined by solving
the multiparticle relativistic equation for two valence electrons~\cite{KotTup87}, with the effective Hamiltonian defined as
$
H_{\mathrm{eff}}(E) = H_{\mathrm{FC}} + \Sigma(E),
$
where $H_{\mathrm{FC}}$ is the Hamiltonian in the frozen-core approximation.
The energy-dependent operator $\Sigma(E)$ which takes into account virtual
core excitations is constructed using second-order perturbation theory in
the CI+MBPT method \cite{DzuFlaKoz96b} and using a linearized coupled-cluster
single-double method in the CI+all-order approach \cite{SafKozJoh09}.
A construction of the effective Hamiltonian in the CI+MBPT and CI+all-order
approximations is described in detail in Refs.~\cite{DzuFlaKoz96b,SafKozJoh09}.

The dynamic polarizability at the imaginary argument is separated into the
valence $\alpha ^{v}(i\omega)$, ionic core $\alpha^{c}(i\omega)$, and $\alpha^{vc}(i\omega )$ parts.
The $vc$ term compensates for the core excitations forbidden by the Pauli principle.
The dominant valence part of the polarizability, $\alpha_k^v(i\omega )$, of
the atomic state $|\Phi \rangle $ can be found by solving the inhomogeneous
equation in the valence space. The wave function of the
intermediate states $|\delta \Phi \rangle $ is expressed as
%===================================================================
%\begin{eqnarray}
%|\delta \Phi \rangle &\equiv& \mathrm{Re}\,\left\{
%\frac{1}{H_{\mathrm{eff}}-E_{\Phi }+i\omega }\,\sum_{i}|\Phi_{i}\rangle
%\langle \Phi _{i}|(T^k_{0})_{\mathrm{eff}}|\Phi \rangle \right\}  \nonumber \\
%&=&\mathrm{Re}\,\left\{ \frac{1}{H_{\mathrm{eff}}-E_{\Phi }+i\omega }%
%\,(T^k_{0})_{\mathrm{eff}}|\Phi \rangle \right\} ,
%\label{delPhi}
%\end{eqnarray}%
$$
|\delta \Phi \rangle
=\mathrm{Re}\,\left\{ \frac{1}{H_{\mathrm{eff}}-E_{\Phi }+i\omega }%
\,(T^k_{0})_{\mathrm{eff}}|\Phi \rangle \right\} ,
\label{delPhi}
$$
where $T^k_{0}$ is the 0-component of the tensor $T$ of multipolarity $k$ and
label ``Re'' means the real part.
Then, $\alpha_k^v(i\omega )$ is given by
%===================================================================
\begin{equation}
\alpha_k^{v}(i\omega )=2\,\langle \Phi |(T^k_0)_{\mathrm{eff} }|\delta \Phi \rangle \,.
\label{alphav}
\end{equation}%
The effective dipole operator $T^k_{\mathrm{eff}}$ includes dominant random-phase
approximation (RPA) corrections.
For the electric-dipole and electric-quadrupole operators, ${\bf d}_0 = T^1_0$ and $Q_0 = T^2_0$.
The ionic core $\alpha^c$ and $\alpha^{vc}$ terms are evaluated in the RPA approximation.

%=================================
%\paragraph{Results and discussion}
% ================================

We start with the calculation of the Yb electric-dipole $\alpha_1$ and electric-quadrupole $\alpha_2$ static   polarizabilities for the $6s^2\,^1\!S_0$, $6s6p\,^3\!P^o_0$, and $6s6p\,^3\!P^o_1$ states
as this allows us to evaluate the accuracy of our approach. The results are summarized in
~\tref{alpha_l}; both scalar, $\alpha_{1s}$, and tensor, $\alpha_{1t}$, parts of the electric dipole polarizability are given for the
$^3\!P^o_1$ state. Higher-order contributions, defined as relative differences of the CI+all-order and CI+MBPT values, are listed in
column labeled ``HO'' in \%.
%###################################################################################
\begin{table}
\caption{The $6s^2\,^1\!S_0$, $6s6p\,^3\!P^o_0$, and $6s6p\,^3\!P^o_1$ electric-dipole,
$\alpha_1$, and electric-quadrupole, $\alpha_2$, static polarizabilities of Yb in
CI+MBPT and CI+all-order approximations (in a.u.).
$\alpha_{ks}$ designates the scalar electric $2^k$-pole
polarizability, and $\alpha_{1t}$ is the tensor part of the dipole polarizability. The (rounded) CI+all-order values are taken as final.
Higher-order contributions, defined as relative differences of the CI+all-order and CI+MBPT values, are listed in
column labeled ``HO'' in \%. The uncertainties are given in parentheses. $^{\rm a}$Ref.~\cite{SafPorCla12}}
\label{alpha_l}
\begin{ruledtabular}
\begin{tabular}{ccccc}
 & \multicolumn{1}{c}{CI+MBPT}  &\multicolumn{1}{c}{CI+All}  &\multicolumn{1}{c}{HO}
 &\multicolumn{1}{c}{Final} \\
\hline
$\alpha_1(^1\!S_0)$    &  138.3  &  140.9  &1.8\%  & 141(2)$^{\rm a}$  \\
$\alpha_1(^3\!P_0^o)$  &  305.9  &  293.2  &-4.3\% &  293(10)$^{\rm a}$ \\
$\alpha_{1s}(^3\!P_1^o)$       &  329.4  &  315.3  &-4.5\% &  315(11) \\
$\alpha_{1t}(^3\!P_1^o)$       &   26.1  &   23.4  &-11.5\%&    23.4(2.7)  \\[0.3pc]
$\alpha_2(^1\!S_0)$            &  2484   &  2559   &2.9\% & 2560(80)  \\
$\alpha_2(^3\!P_0^o)$          & 21294   & 20601   &-3.4\%& 20600(700)  \\
$\alpha_{2s}(^3\!P_1^o)$       & 22923   & 22017   &-4.1\%& 22000(900) \\
\end{tabular}
\end{ruledtabular}
\end{table}
%###################################################################################

In Ref.~\cite{SafPorCla12}, the uncertainties of the  electric-dipole $^1\!S_0$ and $^3\!P^o_0$ polarizabilities were
determined to be 1.4\% and 3.5\%, respectively, based on comparison of the theoretical and experimental results for the $^1\!S_0-\,^3\!P^o_0$
dc Stark shift, magic wavelengths, and the ground-state $C_6$ coefficient for the Yb--Yb dimer.
Table~\ref{alpha_l} illustrates that these uncertainties are slightly lower than the higher-order contributions for these states.
Therefore the difference of the CI+MBPT and CI+all-order results gives a good estimate of the uncertainty.
The calculation of the scalar $^3\!P^o_1$ polarizability is
very similar to the calculation of the $^3\!P^o_0$ polarizability. Thus, the similar calculation accuracy, scaled with the size of the higher orders,
(3.6\%) is assumed. The accuracy of the tensor part of the $^3\!P^o_1$ polarizability
was estimated as the difference between the CI+MBPT and CI+all-order results.
%###################################################################################
\begin{table}[tbp]
\caption{The final (CI+all-order) values of $C_6(\Omega)$ and $C_8(\Omega_{u\!/g})$ coefficients for the Yb--Yb dimers in a.u.
The higher-order corrections, estimated as the relative difference of the CI+all-order and CI+MBPT results, are given in column
labeled ``HO'' in \%. The uncertainties are given in parentheses. The present values are compared with other results
where available.}
\label{C_68}%
\begin{ruledtabular}
\begin{tabular}{cccll}
                      &                      &    HO    &   Final                 & Other \\
\hline
$^1\!S_0 +\, ^1\!S_0$     & $C_6^{\rm ~a}$   &   1.5\%  &  1929(39)               & 1932(35)$^{\rm b}$ \\
                          & $C_8$            &   2.9\%  &  1.88(6)$\times 10^5$   & 1.9(5)$\times 10^5$\,$^{\rm b}$\\[0.3pc]
$^1\!S_0 +\, ^3\!P_0^o$   & $C_6$            &  -1.9\%  &  2561(95)               & 2709(338)$^{\rm c}$ \\[0.3pc]
$^3\!P_0^o +\, ^3\!P_0^o$ & $C_6$            &  -4.5\%  &  3746(170)              & 3886(360)$^{\rm c}$ \\[0.3pc]
$^1\!S_0 +\, ^3\!P_1^o$   & $C_6(0)$ &  -0.3\%  &  2640(100)              & 2410(220)$^{\rm d}$ \\
                          & $C_6(1)$ &  -1.4\%  &  2785(110)              & 2283.6$^{\rm e}$ \\[0.2pc]
                          & $C_8(0_{u\!/g})$ &  -0.3\%  &  3.19(14)$\times 10^5$  &                   \\
                          & $C_8(1_{u\!/g})$ &  -0.4\%  &  4.11(18)$\times 10^5$  &                   \\
\end{tabular}
\end{ruledtabular}\begin{flushleft}
$^{\rm a}${Ref.~\cite{SafPorCla12}, theory.}
$^{\rm b}${Ref.~\cite{KitEnoKas08}, experiment.}
$^{\rm c}${Ref.~\cite{DzuDer10}, theory.}
$^{\rm d}${Ref.~\cite{BorCiuJul09}, experiment; the error includes only uncertainty of the fit.}
$^{\rm e}${Ref.~\cite{TakSaiTak12}, experiment.}\end{flushleft}
\end{table}
%###################################################################################

To the best of our knowledge, there are no experimental data for the electric-quadrupole
polarizabilities listed in Table~\ref{alpha_l}, nor for any of  the transitions that give dominant  contributions
 to these polarizabilities, such as the $6s^2 ~^1\!S_0 - 5d6s\, ^1\!D_2$ transition that contributes about 75\% to $\alpha_2(^1\!S_0)$.
Any precise experimental data for the $5d6s\, ^1\!D_2$ state (lifetime, oscillator strengths, etc)
would provide benchmarks relevant to the accuracies of the quadrupole polarizabilities.
We estimated the uncertainties of these values (3-4\%)
as the differences between the results obtained in the CI+MBPT and CI+all-order approximations.
We note that the higher-order corrections contribute with an opposite sign to the $^1\!S_0$ and all $^3\!P^o_J$ polarizabilities, which will affect the
uncertainties of their combined properties.

The values of the $C_{3}$ coefficients  obtained in the CI+MBPT and
CI+all-order approximations for the Yb--Yb $^1\!S_0 +\, ^3\!P_1^o$ dimer
are presented in Table~III of the Supplemental Material~\cite{SupMat}. These coefficients depend
entirely on the value of the $\langle 6s6p\,^3\!P_1^o||d||6s^2\,^1\!S_0\rangle$ matrix element
which has been discussed in ~\cite{SafPorCla12}.

To determine the $C_6$ and $C_8$ coefficients, we calculate
frequency-dependent electric-dipole
and quadrupole polarizabilities at imaginary frequencies for the $^1\!S_0$ and $^3\!P_{0,1}^o$ states.
The $\omega=0$ values correspond to static polarizabilities discussed above.
The integrals over $\omega$ needed for evaluation of the $C_6$ and $C_8$ coefficients are calculated using
Gaussian quadrature of the integrand computed on the finite grid of discrete imaginary frequencies
~\cite{BisPip92}.
Details of the calculations, including the contributions of various terms to polarizabilities and the van der Waals coefficients, will be discussed in a
subsequent paper.
The results are listed in Table~\ref{C_68}. We note that the $C_6$ coefficients depend on
$\Omega$ but do not depend on $g/u$ symmetry. While there are small differences in the values of $C_8$ gerade/ungerade coefficients,
these are less then our estimated uncertainties so we do not list separate values.

To estimate the uncertainties of the $C_6$ and $C_8$ coefficients, we compare frequency-dependent
polarizabilities calculated in the CI+MBPT and CI+all-order approximations for all $\omega$ used in our finite grid.
We find that the difference between the CI+all-order and CI+MBPT frequency-dependent polarizability values is the largest
for $\omega=0$ and decreases significantly with increasing $\omega$. Therefore, the upper bound on the relative
uncertainties of the terms in the expressions of the $C_6$ and $C_8$ coefficients that contain integrals over $\omega$
may be estimated  by the uncertainties in the corresponding static polarizabilities of atoms $A$ and $B$ (listed in Table~\ref{alpha_l}) added in quadratures.
The uncertainties of the terms that do not contain integrals over $\omega$ can be obtained using the uncertainties of the contributing matrix elements or
using the difference of the CI+MBPT and CI+all-order values of the entire term.
As a result, we estimated the fractional uncertainties of the $C_6$ coefficients for the
$^1\!S_0 +\, ^3\!P^o_{0,1}$  dimer at the level of 4\%. The uncertainty of the $C_8\, (^1\!S_0 +\, ^1\!S_0)$ coefficient
is 3.2\% and the uncertainty of the $C_8\,(^1\!S_0 +\, ^3\!P^o_1)$ coefficients is 4.5\%.
The difference of the CI+all-order and CI+MBPT results (4.5\%) is taken as an uncertainty for
the $C_6\, (^3\!P^o_0+\,^3\!P^o_0)$ coefficient.
%###################################################################################
% ==================================================================================
%\paragraph{$C_{6}$ coefficients for the Yb--Rb dimers}
\label{YbRb}
% ==================================================================================

We have also investigated the molecular potentials asymptotically connecting to
the $6s6p~^{3}\!P_{1}^{o}+5s~^2\!S_{1/2}$ and $6s^2~^{1}\!S_{0}+5p~^2\!P^o_{1/2}$ atomic states
of the Yb--Rb dimer since these cases are of particular experimental interest.
We use shorten, $5s$ and $5p_{1/2}$, notations for the Rb states below.
The ground state case $^{1}\!S_{0}+5s$ was previously considered in~\cite{SafPorCla12} and the $C_6$ coefficient was obtained.
The $C_8$ coefficient for this case is calculated in the present work.

 We assume that the coupling scheme can be described
by the usual Hund's case (c), i.e., $\Omega$ is a good quantum number for all Yb--Rb dimers studied here.
The expression for the $C_{6}\, (^{3}\!P_{1}^{o}+5s)$ coefficient is given in the Supplemental Material~\cite{SupMat}.
It can be shown that the $C_6$ coefficient for the Yb--Rb $^{1}\!S_{0}+5p_{1/2}$  dimer is given by
%--------------------------------------------------------------------------------
\begin{eqnarray}
C_6 = \frac{3}{\pi} \int_0^\infty \alpha^A_1 (i\omega)\, \alpha^B_1(i\omega )\, d\omega
+ D^2 \, \alpha^B_1(\omega_{As}) ,
\label{C6_5p1S0}
\end{eqnarray}%
%--------------------------------------------------------------------------------
where  $D=|\langle 5p_{1/2} ||d|| 5s \rangle|$, $\alpha^A_1(i\omega)$ and  $\alpha^B_1(i\omega)$
are the dynamic electric-dipole polarizabilities of the Rb
$5p_{1/2}$ state and the Yb ground state
at the imaginary argument, and $\alpha^B_1(\omega_{As})$ is the  frequency-dependent polarizability of
the Yb ground state at $\omega_{As} \equiv E_{5p_{1/2}}-E_{5s}$.
\begin{table}[tbp]
\caption{ The final  values of the $C_6$ and $C_8$ coefficients for the Yb--Rb dimers in a.u. Rb
electric dipole ($\alpha_1$) and quadrupole ($\alpha_2$) static polarizabilities are listed for reference.
A comparison is given with other theory~\cite{AroSafCla07,BruHut13} and experiments~\cite{MilKraHun94,NemBauMun09}.
The uncertainties are given in parentheses.}
\label{alphRb}%
\begin{ruledtabular}
\begin{tabular}{llcr}
       &                                   & \multicolumn{1}{c}{Present} &\multicolumn{1}{c}{Other} \\
\hline
Rb    &$\alpha_1(5s)$                      & 318.6(6)$^a$                & 318.4(6)~\cite{AroSafCla07}\\
Rb    &$\alpha_2(5s)$                      & 6520(80)$^a$                & 6525(37)~\cite{AroSafCla07}\\
Rb    &$\alpha_1(5p_{1/2})$                & 810.1(8)$^b$                & 811(6)~\cite{MilKraHun94}\\
Yb--Rb&$C_6(^1\!S_0+5s)$                   & 2837(57)$^c$                & 2830~\cite{BruHut13}  \\
Yb--Rb&$C_8(^1\!S_0+5s)$                   & 3.20(7)$\times 10^5$        & \\
Yb--Rb&$C_6(^1\!S_0+5p_{1/2})$             & 7610(115)                   & 5684(98)$^d$~\cite{NemBauMun09}\\
Yb--Rb&$C_6(^3\!P^o_1+5s)$ ~$\Omega=1/2$   & 3955(160)                   & \\
Yb--Rb&$C_6(^3\!P^o_1+5s)$ ~$\Omega=3/2$   & 4470(180)                   & \\
\end{tabular}
\end{ruledtabular}\begin{flushleft}
$^a$Refs.~\cite{PorDer03,DerPorBab10}.
$^b$Ref.~\cite{ZhuDalPor04}.
$^c$Ref.~\cite{SafPorCla12}.
$^d$This uncertainty includes only error of the fit with Leroy-Bernstein method.
\end{flushleft}
\end{table}

The ground state Rb-Rb $C_6$  and $C_8$ calculations were previously carried out in \cite{PorDer03,DerPorBab10}, and we use the $5s$ dynamic polarizability
at imaginary frequencies calculated in those works.
The Rb $5p_{1/2}$ polarizability was calculated in \cite{ZhuDalPor04}; its uncertainty was
estimated at the level of 0.1\%.
The values of Rb dipole ($\alpha_1$) and quadrupole ($\alpha_2$) static polarizabilities
 used in this work are given  in Table~\ref{alphRb}.
We evaluate the $C_6(\Omega)$ coefficients for the Yb--Rb $^3\!P_1^o+\,5s$ and $^{1}\!S_{0}+5p_{1/2}$ dimers
using the Rb dynamic polarizabilities from Refs.~\cite{DerPorBab10,ZhuDalPor04}
and Yb polarizabilities at imaginary frequencies used in our calculation of the long-range interaction coefficients discussed above for the Yb--Yb dimers.

 Since the uncertainties of the Rb $5s$ and $5p_{1/2}$ static polarizabilities are negligible in comparison to the uncertainties
of the Yb scalar $^3\!P^o_1$ and $^1\!S_0$ static  polarizabilities, the latter determine the uncertainties of the $C_6$ coefficients.
Our final results are given in Table~\ref{alphRb}.

The $C_8$ coefficient for the Yb--Rb $^{1}\!S_{0}+5s$ dimer is calculated using Eq.~(\ref{vdW}). The final value is listed
in Table~\ref{alphRb}. The uncertainty is determined from the uncertainties of the ground state dipole and  quadrupole polarizabilities for Yb and Rb.

To provide additional verification of this value and its accuracy, we derived a semi-empirical formula for the $C_8^{AB}$ coefficient
of heteronuclear dimers, when both atoms are in spherically symmetric states,
following a method suggested by Tang~\cite{Tan69}.
The resulting expression involves static dipole $\alpha_1(0)$ and quadrupole $\alpha_2(0)$ polarizabilities
of the atomic states $A$ and $B$, and the $C_6$ and $C_8$ coefficients for the homonucler dimers:
    \begin{equation}
C_8^{AB} \approx \frac{15}{4} \left( \frac{\alpha_1^A(0)\,\alpha_2^B(0)}{\phi_1^A + \phi_2^B} +
 \frac{\alpha_2^A(0)\,\alpha_1^B(0)}{\phi_2^A + \phi_1^B} \right).
\label{C8approx}
\end{equation}
The quantities $\phi_1^X$ and $\phi_2^X$ for an atom $X$ are given by
%\begin{eqnarray}
%\phi_1^X&= & \frac{3}{4}\,\frac{(\alpha_1^X(0))^2}{C_6^{XX}}, \nonumber \\
%\phi_2^X &=& \frac{15}{2}\, \frac{\alpha_1^X(0)\,\alpha_2^X(0)}{C_8^{XX}} - \phi_1^X.
%\end{eqnarray}
%\label{phi1}
$$
\phi_1^X=  \frac{3}{4}\,\frac{(\alpha_1^X(0))^2}{C_6^{XX}}, ~~~
\phi_2^X = \frac{15}{2}\, \frac{\alpha_1^X(0)\,\alpha_2^X(0)}{C_8^{XX}} - \phi_1^X.
$$
Using this formula, we obtained $C_8 \approx 3.20 \times 10^5$~a.u. for the Yb--Rb $^1\!S_0 + 5s$
dimer which is identical to our value obtained with use of Eq.~(\ref{vdW}) and given in Table~\ref{alphRb}.

To conclude, we evaluated the $C_6$ and $C_8$ coefficients for the Yb--Yb and Yb--Rb dimers of particular interest
for studying the quantum gas mixtures. The uncertainties of all properties are determined to allow future benchmark tests of molecular theory and experiment.
 Most of these properties are determined for the first time.  Methodology developed in this
work can be used to evaluate properties of other dimers with excited atoms that have a strong decay channel.

We thank P. Julienne and T. Porto for helpful discussions.
This research was performed under the sponsorship of the
U.S. Department of Commerce, National Institute of Standards and
Technology, and was supported by the National Science Foundation
under Physics Frontiers Center Grant No. PHY-0822671 and by the
Office of Naval Research. The work of S.G.P. was supported in part by
US NSF Grant No. PHY-1212442 and RFBR Grant No.\ 11-02-00943. Work of A.D. was supported in part by the US NSF Grant No. PHY-0969580.
%===================================================================

%\bibliographystyle{revtex}
%\bibliography{YbC6} %library_apd

\begin{thebibliography}{39}
\expandafter\ifx\csname natexlab\endcsname\relax\def\natexlab#1{#1}\fi
\expandafter\ifx\csname bibnamefont\endcsname\relax
  \def\bibnamefont#1{#1}\fi
\expandafter\ifx\csname bibfnamefont\endcsname\relax
  \def\bibfnamefont#1{#1}\fi
\expandafter\ifx\csname citenamefont\endcsname\relax
  \def\citenamefont#1{#1}\fi
\expandafter\ifx\csname url\endcsname\relax
  \def\url#1{\texttt{#1}}\fi
\expandafter\ifx\csname urlprefix\endcsname\relax\def\urlprefix{URL }\fi
\providecommand{\bibinfo}[2]{#2}
\providecommand{\eprint}[2][]{\url{#2}}

\bibitem[{\citenamefont{Takasu et~al.}(2004)}]{TakKomHon04}
\bibinfo{author}{\bibfnamefont{Y.}~\bibnamefont{Takasu}} \bibnamefont{et~al.},
  \bibinfo{journal}{Phys. Rev. Lett.} \textbf{\bibinfo{volume}{93}},
  \bibinfo{pages}{123202} (\bibinfo{year}{2004}).

\bibitem[{\citenamefont{{Lemke} et~al.}(2009)}]{Ybclock}
\bibinfo{author}{\bibfnamefont{N.~D.} \bibnamefont{{Lemke}}}
  \bibnamefont{et~al.}, \bibinfo{journal}{Phys. Rev. Lett.}
  \textbf{\bibinfo{volume}{103}}, \bibinfo{eid}{063001} (\bibinfo{year}{2009}).

\bibitem[{\citenamefont{Gorshkov et~al.}(2009)}]{GorReyDal09}
\bibinfo{author}{\bibfnamefont{A.~V.} \bibnamefont{Gorshkov}}
  \bibnamefont{et~al.}, \bibinfo{journal}{Phys. Rev. Lett.}
  \textbf{\bibinfo{volume}{102}}, \bibinfo{pages}{110503}
  (\bibinfo{year}{2009}).

\bibitem[{\citenamefont{{Tsigutkin} et~al.}(2009)}]{PNC1}
\bibinfo{author}{\bibfnamefont{K.}~\bibnamefont{{Tsigutkin}}}
  \bibnamefont{et~al.}, \bibinfo{journal}{Phys. Rev. Lett.}
  \textbf{\bibinfo{volume}{103}}, \bibinfo{eid}{071601} (\bibinfo{year}{2009}).

\bibitem[{\citenamefont{Borkowski et~al.}(2009)}]{BorCiuJul09}
\bibinfo{author}{\bibfnamefont{M.}~\bibnamefont{Borkowski}}
  \bibnamefont{et~al.}, \bibinfo{journal}{Phys. Rev. A}
  \textbf{\bibinfo{volume}{80}}, \bibinfo{pages}{012715}
  (\bibinfo{year}{2009}).

\bibitem[{\citenamefont{Tojo et~al.}(2006)}]{TojKitEno06}
\bibinfo{author}{\bibfnamefont{S.}~\bibnamefont{Tojo}} \bibnamefont{et~al.},
  \bibinfo{journal}{Phys. Rev. Lett.} \textbf{\bibinfo{volume}{96}},
  \bibinfo{pages}{153201} (\bibinfo{year}{2006}).

\bibitem[{\citenamefont{Enomoto et~al.}(2008)\citenamefont{Enomoto, Kitagawa,
  Tojo, and Takahashi}}]{EnoKitToj08}
\bibinfo{author}{\bibfnamefont{K.}~\bibnamefont{Enomoto}},
  \bibinfo{author}{\bibfnamefont{M.}~\bibnamefont{Kitagawa}},
  \bibinfo{author}{\bibfnamefont{S.}~\bibnamefont{Tojo}}, \bibnamefont{and}
  \bibinfo{author}{\bibfnamefont{Y.}~\bibnamefont{Takahashi}},
  \bibinfo{journal}{Phys. Rev. Lett.} \textbf{\bibinfo{volume}{100}},
  \bibinfo{pages}{123001} (\bibinfo{year}{2008}).

\bibitem[{\citenamefont{Kitagawa et~al.}(2008)}]{KitEnoKas08}
\bibinfo{author}{\bibfnamefont{M.}~\bibnamefont{Kitagawa}}
  \bibnamefont{et~al.}, \bibinfo{journal}{Phys. Rev. A}
  \textbf{\bibinfo{volume}{77}}, \bibinfo{pages}{012719}
  (\bibinfo{year}{2008}).

\bibitem[{\citenamefont{{Yamazaki} et~al.}(2013)}]{YamTaiSug13}
\bibinfo{author}{\bibfnamefont{R.}~\bibnamefont{{Yamazaki}}}
  \bibnamefont{et~al.}, \bibinfo{journal}{\pra} \textbf{\bibinfo{volume}{87}},
  \bibinfo{eid}{010704} (\bibinfo{year}{2013}).

\bibitem[{\citenamefont{Takasu et~al.}(2012)}]{TakSaiTak12}
\bibinfo{author}{\bibfnamefont{Y.}~\bibnamefont{Takasu}} \bibnamefont{et~al.},
  \bibinfo{journal}{Phys. Rev. Lett.} \textbf{\bibinfo{volume}{108}},
  \bibinfo{pages}{173002} (\bibinfo{year}{2012}).

\bibitem[{\citenamefont{{Nemitz} et~al.}(2009)}]{NemBauMun09}
\bibinfo{author}{\bibfnamefont{N.}~\bibnamefont{{Nemitz}}}
  \bibnamefont{et~al.}, \bibinfo{journal}{Phys. Rev. A}
  \textbf{\bibinfo{volume}{79}}, \bibinfo{eid}{061403} (\bibinfo{year}{2009}).

\bibitem[{\citenamefont{M\"{u}nchow et~al.}(2011)\citenamefont{M\"{u}nchow,
  Bruni, Madalinski, and G\"{o}rlitz}}]{MunBruMad11}
\bibinfo{author}{\bibfnamefont{F.}~\bibnamefont{M\"{u}nchow}},
  \bibinfo{author}{\bibfnamefont{C.}~\bibnamefont{Bruni}},
  \bibinfo{author}{\bibfnamefont{M.}~\bibnamefont{Madalinski}},
  \bibnamefont{and}
  \bibinfo{author}{\bibfnamefont{A.}~\bibnamefont{G\"{o}rlitz}},
  \bibinfo{journal}{Phys. Chem. Chem. Phys.} \textbf{\bibinfo{volume}{13}},
  \bibinfo{pages}{18734} (\bibinfo{year}{2011}).

\bibitem[{\citenamefont{{Baumer} et~al.}(2011)}]{BauMunGor11}
\bibinfo{author}{\bibfnamefont{F.}~\bibnamefont{{Baumer}}}
  \bibnamefont{et~al.}, \bibinfo{journal}{\pra} \textbf{\bibinfo{volume}{83}},
  \bibinfo{eid}{040702} (\bibinfo{year}{2011}).

\bibitem[{\citenamefont{Hudson et~al.}(2011)}]{HudKarSma11}
\bibinfo{author}{\bibfnamefont{J.~J.} \bibnamefont{Hudson}}
  \bibnamefont{et~al.}, \bibinfo{journal}{Nature}
  \textbf{\bibinfo{volume}{473}}, \bibinfo{pages}{493} (\bibinfo{year}{2011}).

\bibitem[{\citenamefont{Micheli et~al.}(2006)\citenamefont{Micheli, Brennen,
  and Zoller}}]{MicBreZol06}
\bibinfo{author}{\bibfnamefont{A.}~\bibnamefont{Micheli}},
  \bibinfo{author}{\bibfnamefont{G.~K.} \bibnamefont{Brennen}},
  \bibnamefont{and} \bibinfo{author}{\bibfnamefont{P.}~\bibnamefont{Zoller}},
  \bibinfo{journal}{Nature Phys.} \textbf{\bibinfo{volume}{2}},
  \bibinfo{pages}{341} (\bibinfo{year}{2006}).

\bibitem[{\citenamefont{{Reichenbach} et~al.}(2009)\citenamefont{{Reichenbach},
  {Julienne}, and {Deutsch}}}]{ReiJulDeu09}
\bibinfo{author}{\bibfnamefont{I.}~\bibnamefont{{Reichenbach}}},
  \bibinfo{author}{\bibfnamefont{P.~S.} \bibnamefont{{Julienne}}},
  \bibnamefont{and} \bibinfo{author}{\bibfnamefont{I.~H.}
  \bibnamefont{{Deutsch}}}, \bibinfo{journal}{\pra}
  \textbf{\bibinfo{volume}{80}}, \bibinfo{eid}{020701} (\bibinfo{year}{2009}).

\bibitem[{\citenamefont{{Meyer} and {Bohn}}(2009)}]{MeyBoh09}
\bibinfo{author}{\bibfnamefont{E.~R.} \bibnamefont{{Meyer}}} \bibnamefont{and}
  \bibinfo{author}{\bibfnamefont{J.~L.} \bibnamefont{{Bohn}}},
  \bibinfo{journal}{\pra} \textbf{\bibinfo{volume}{80}}, \bibinfo{eid}{042508}
  (\bibinfo{year}{2009}).

\bibitem[{\citenamefont{{Sanders} et~al.}(2011)\citenamefont{{Sanders},
  {Odong}, {Javanainen}, and {Mackie}}}]{SanOdoJav11}
\bibinfo{author}{\bibfnamefont{J.~C.} \bibnamefont{{Sanders}}},
  \bibinfo{author}{\bibfnamefont{O.}~\bibnamefont{{Odong}}},
  \bibinfo{author}{\bibfnamefont{J.}~\bibnamefont{{Javanainen}}},
  \bibnamefont{and} \bibinfo{author}{\bibfnamefont{M.}~\bibnamefont{{Mackie}}},
  \bibinfo{journal}{\pra} \textbf{\bibinfo{volume}{83}}, \bibinfo{eid}{031607}
  (\bibinfo{year}{2011}).

\bibitem[{\citenamefont{Ludlow et~al.}(2011)\citenamefont{Ludlow, Lemke,
  Sherman, Oates, Qu\'{e}m\'{e}ner, von Stecher, and Rey}}]{LudLemShe11}
\bibinfo{author}{\bibfnamefont{A.~D.} \bibnamefont{Ludlow}},
  \bibinfo{author}{\bibfnamefont{N.~D.} \bibnamefont{Lemke}},
  \bibinfo{author}{\bibfnamefont{J.~A.} \bibnamefont{Sherman}},
  \bibinfo{author}{\bibfnamefont{C.~W.} \bibnamefont{Oates}},
  \bibinfo{author}{\bibfnamefont{G.}~\bibnamefont{Qu\'{e}m\'{e}ner}},
  \bibinfo{author}{\bibfnamefont{J.}~\bibnamefont{von Stecher}},
  \bibnamefont{and} \bibinfo{author}{\bibfnamefont{A.~M.} \bibnamefont{Rey}},
  \bibinfo{journal}{Phys. Rev. A} \textbf{\bibinfo{volume}{84}},
  \bibinfo{pages}{52724} (\bibinfo{year}{2011}).

\bibitem[{\citenamefont{Bishof et~al.}(2011)\citenamefont{Bishof, Lin,
  Swallows, Gorshkov, Ye, and Rey}}]{BisLinSwa11}
\bibinfo{author}{\bibfnamefont{M.}~\bibnamefont{Bishof}},
  \bibinfo{author}{\bibfnamefont{Y.}~\bibnamefont{Lin}},
  \bibinfo{author}{\bibfnamefont{M.~D.} \bibnamefont{Swallows}},
  \bibinfo{author}{\bibfnamefont{a.~V.} \bibnamefont{Gorshkov}},
  \bibinfo{author}{\bibfnamefont{J.}~\bibnamefont{Ye}}, \bibnamefont{and}
  \bibinfo{author}{\bibfnamefont{A.~M.} \bibnamefont{Rey}},
  \bibinfo{journal}{Phys. \ Rev. \ Lett.} \textbf{\bibinfo{volume}{106}},
  \bibinfo{pages}{250801} (\bibinfo{year}{2011}).

\bibitem[{YeR()}]{YeReyScience2013forthcoming}
\bibinfo{note}{{J.} Ye and A.-M. Rey, private communication (2013)}.

\bibitem[{\citenamefont{Ni et~al.}(2008)\citenamefont{Ni, Ospelkaus,
  de~Miranda, Pe`er, Neyenhuis, Zirbel, Kotochigova, Julienne, Jin, and
  Ye}}]{NiOspMir08}
\bibinfo{author}{\bibfnamefont{K.~K.} \bibnamefont{Ni}},
  \bibinfo{author}{\bibfnamefont{S.}~\bibnamefont{Ospelkaus}},
  \bibinfo{author}{\bibfnamefont{M.~H.~G.} \bibnamefont{de~Miranda}},
  \bibinfo{author}{\bibfnamefont{A.}~\bibnamefont{Pe`er}},
  \bibinfo{author}{\bibfnamefont{B.}~\bibnamefont{Neyenhuis}},
  \bibinfo{author}{\bibfnamefont{J.~J.} \bibnamefont{Zirbel}},
  \bibinfo{author}{\bibfnamefont{S.}~\bibnamefont{Kotochigova}},
  \bibinfo{author}{\bibfnamefont{P.~S.} \bibnamefont{Julienne}},
  \bibinfo{author}{\bibfnamefont{D.~S.} \bibnamefont{Jin}}, \bibnamefont{and}
  \bibinfo{author}{\bibfnamefont{J.}~\bibnamefont{Ye}},
  \bibinfo{journal}{Science} \textbf{\bibinfo{volume}{322}},
  \bibinfo{pages}{231} (\bibinfo{year}{2008}).

\bibitem[{\citenamefont{{Safronova} et~al.}(2012)\citenamefont{{Safronova},
  {Porsev}, and {Clark}}}]{SafPorCla12}
\bibinfo{author}{\bibfnamefont{M.~S.} \bibnamefont{{Safronova}}},
  \bibinfo{author}{\bibfnamefont{S.~G.} \bibnamefont{{Porsev}}},
  \bibnamefont{and} \bibinfo{author}{\bibfnamefont{C.~W.}
  \bibnamefont{{Clark}}}, \bibinfo{journal}{Phys. Rev. Lett.}
  \textbf{\bibinfo{volume}{109}}, \bibinfo{eid}{230802} (\bibinfo{year}{2012}).

\bibitem[{\citenamefont{Patil and Tang}(1997)}]{PatTan97}
\bibinfo{author}{\bibfnamefont{S.~H.} \bibnamefont{Patil}} \bibnamefont{and}
  \bibinfo{author}{\bibfnamefont{K.~T.} \bibnamefont{Tang}},
  \bibinfo{journal}{J. Chem.\ Phys.} \textbf{\bibinfo{volume}{106}},
  \bibinfo{pages}{2298} (\bibinfo{year}{1997}).

\bibitem[{Sup()}]{SupMat}
\bibinfo{note}{See Supplemental Material at [URL] for details $C_6$ and $C_8$
  formulas and discussion of the $C_3$ coefficients.}

\bibitem[{\citenamefont{Dzuba et~al.}(1996)\citenamefont{Dzuba, Flambaum, and
  Kozlov}}]{DzuFlaKoz96b}
\bibinfo{author}{\bibfnamefont{V.~A.} \bibnamefont{Dzuba}},
  \bibinfo{author}{\bibfnamefont{V.~V.} \bibnamefont{Flambaum}},
  \bibnamefont{and} \bibinfo{author}{\bibfnamefont{M.~G.}
  \bibnamefont{Kozlov}}, \bibinfo{journal}{Phys.\ Rev.\ A}
  \textbf{\bibinfo{volume}{54}}, \bibinfo{pages}{3948} (\bibinfo{year}{1996}).

\bibitem[{\citenamefont{Kozlov}(2004)}]{Koz04}
\bibinfo{author}{\bibfnamefont{M.~G.} \bibnamefont{Kozlov}},
  \bibinfo{journal}{Int. J. Quant. Chem.} \textbf{\bibinfo{volume}{100}},
  \bibinfo{pages}{336} (\bibinfo{year}{2004}).

\bibitem[{\citenamefont{{Safronova} et~al.}(2009)\citenamefont{{Safronova},
  {Kozlov}, {Johnson}, and {Jiang}}}]{SafKozJoh09}
\bibinfo{author}{\bibfnamefont{M.~S.} \bibnamefont{{Safronova}}},
  \bibinfo{author}{\bibfnamefont{M.~G.} \bibnamefont{{Kozlov}}},
  \bibinfo{author}{\bibfnamefont{W.~R.} \bibnamefont{{Johnson}}},
  \bibnamefont{and} \bibinfo{author}{\bibfnamefont{D.}~\bibnamefont{{Jiang}}},
  \bibinfo{journal}{Phys. Rev. A} \textbf{\bibinfo{volume}{80}},
  \bibinfo{pages}{012516} (\bibinfo{year}{2009}).

\bibitem[{\citenamefont{{Safronova} et~al.}(2011)\citenamefont{{Safronova},
  {Kozlov}, and {Clark}}}]{SafKozCla11}
\bibinfo{author}{\bibfnamefont{M.~S.} \bibnamefont{{Safronova}}},
  \bibinfo{author}{\bibfnamefont{M.~G.} \bibnamefont{{Kozlov}}},
  \bibnamefont{and} \bibinfo{author}{\bibfnamefont{C.~W.}
  \bibnamefont{{Clark}}}, \bibinfo{journal}{Phys. Rev. Lett.}
  \textbf{\bibinfo{volume}{107}}, \bibinfo{pages}{143006}
  (\bibinfo{year}{2011}).

\bibitem[{\citenamefont{Kotochigova and Tupitsyn}(1987)}]{KotTup87}
\bibinfo{author}{\bibfnamefont{S.~A.} \bibnamefont{Kotochigova}}
  \bibnamefont{and} \bibinfo{author}{\bibfnamefont{I.~I.}
  \bibnamefont{Tupitsyn}}, \bibinfo{journal}{J. Phys. B}
  \textbf{\bibinfo{volume}{20}}, \bibinfo{pages}{4759} (\bibinfo{year}{1987}).

\bibitem[{\citenamefont{Dzuba and Derevianko}(2010)}]{DzuDer10}
\bibinfo{author}{\bibfnamefont{V.~A.} \bibnamefont{Dzuba}} \bibnamefont{and}
  \bibinfo{author}{\bibfnamefont{A.}~\bibnamefont{Derevianko}},
  \bibinfo{journal}{J. Phys. B} \textbf{\bibinfo{volume}{43}},
  \bibinfo{pages}{074011} (\bibinfo{year}{2010}).

\bibitem[{\citenamefont{Bishop and Pipin}(1992)}]{BisPip92}
\bibinfo{author}{\bibfnamefont{D.~M.} \bibnamefont{Bishop}} \bibnamefont{and}
  \bibinfo{author}{\bibfnamefont{J.}~\bibnamefont{Pipin}}, \bibinfo{journal}{J.
  Chem. Phys.} \textbf{\bibinfo{volume}{97}}, \bibinfo{pages}{3375}
  (\bibinfo{year}{1992}).

\bibitem[{\citenamefont{Arora et~al.}(2007)\citenamefont{Arora, Safronova, and
  Clark}}]{AroSafCla07}
\bibinfo{author}{\bibfnamefont{B.}~\bibnamefont{Arora}},
  \bibinfo{author}{\bibfnamefont{M.~S.} \bibnamefont{Safronova}},
  \bibnamefont{and} \bibinfo{author}{\bibfnamefont{C.~W.} \bibnamefont{Clark}},
  \bibinfo{journal}{Phys. Rev. A} \textbf{\bibinfo{volume}{76}},
  \bibinfo{pages}{052516} (\bibinfo{year}{2007}).

\bibitem[{\citenamefont{Brue and Hutson}(2013)}]{BruHut13}
\bibinfo{author}{\bibfnamefont{D.~A.} \bibnamefont{Brue}} \bibnamefont{and}
  \bibinfo{author}{\bibfnamefont{J.~M.} \bibnamefont{Hutson}},
  \bibinfo{journal}{Phys. Rev. A} \textbf{\bibinfo{volume}{87}},
  \bibinfo{pages}{052709} (\bibinfo{year}{2013}).

\bibitem[{\citenamefont{{Miller} et~al.}(1994)\citenamefont{{Miller}, {Krause},
  and {Hunter}}}]{MilKraHun94}
\bibinfo{author}{\bibfnamefont{K.~E.} \bibnamefont{{Miller}}},
  \bibinfo{author}{\bibfnamefont{D.}~\bibnamefont{{Krause}},
  \bibfnamefont{Jr.}}, \bibnamefont{and} \bibinfo{author}{\bibfnamefont{L.~R.}
  \bibnamefont{{Hunter}}}, \bibinfo{journal}{\pra}
  \textbf{\bibinfo{volume}{49}}, \bibinfo{pages}{5128} (\bibinfo{year}{1994}).

\bibitem[{\citenamefont{Porsev and Derevianko}(2003)}]{PorDer03}
\bibinfo{author}{\bibfnamefont{S.~G.} \bibnamefont{Porsev}} \bibnamefont{and}
  \bibinfo{author}{\bibfnamefont{A.}~\bibnamefont{Derevianko}},
  \bibinfo{journal}{J. Chem. Phys.} \textbf{\bibinfo{volume}{119}},
  \bibinfo{pages}{844} (\bibinfo{year}{2003}).

\bibitem[{\citenamefont{Derevianko et~al.}(2010)\citenamefont{Derevianko,
  Porsev, and Babb}}]{DerPorBab10}
\bibinfo{author}{\bibfnamefont{A.}~\bibnamefont{Derevianko}},
  \bibinfo{author}{\bibfnamefont{S.~G.} \bibnamefont{Porsev}},
  \bibnamefont{and} \bibinfo{author}{\bibfnamefont{J.~F.} \bibnamefont{Babb}},
  \bibinfo{journal}{At. Data Nucl. Data Tables} \textbf{\bibinfo{volume}{96}},
  \bibinfo{pages}{323} (\bibinfo{year}{2010}).

\bibitem[{\citenamefont{Zhu et~al.}(2004)\citenamefont{Zhu, Dalgarno, Porsev,
  and Derevianko}}]{ZhuDalPor04}
\bibinfo{author}{\bibfnamefont{C.}~\bibnamefont{Zhu}},
  \bibinfo{author}{\bibfnamefont{A.}~\bibnamefont{Dalgarno}},
  \bibinfo{author}{\bibfnamefont{S.~G.} \bibnamefont{Porsev}},
  \bibnamefont{and}
  \bibinfo{author}{\bibfnamefont{A.}~\bibnamefont{Derevianko}},
  \bibinfo{journal}{Phys. Rev. A} \textbf{\bibinfo{volume}{70}},
  \bibinfo{pages}{032722} (\bibinfo{year}{2004}).

\bibitem[{\citenamefont{Tang}(1969)}]{Tan69}
\bibinfo{author}{\bibfnamefont{K.~T.} \bibnamefont{Tang}},
  \bibinfo{journal}{Phys. Rev.} \textbf{\bibinfo{volume}{177}},
  \bibinfo{pages}{108} (\bibinfo{year}{1969}).

\end{thebibliography}

\end{document}